%% file: ms.tex
\documentclass[sigconf]{acmart}
\pdfoutput=1

\usepackage{enumitem}
\usepackage{bm}
\usepackage{multirow}
\usepackage{subfigure}

\usepackage{amsmath}
\usepackage{cases}

\usepackage{balance}

\usepackage[titletoc]{appendix}

\usepackage[]{fixltx2e}
\usepackage[bottom]{footmisc}

\AtBeginDocument{%
  \providecommand\BibTeX{{%
    \normalfont B\kern-0.5em{\scshape i\kern-0.25em b}\kern-0.8em\TeX}}}

\copyrightyear{2022}
\acmYear{2022}
\setcopyright{acmcopyright}\acmConference[WWW '22]{Proceedings of the ACM Web Conference 2022}{April 25--29, 2022}{Virtual Event, Lyon, France}
\acmBooktitle{Proceedings of the ACM Web Conference 2022 (WWW '22), April 25--29, 2022, Virtual Event, Lyon, France}
\acmPrice{15.00}
\acmDOI{10.1145/3485447.3512098}
\acmISBN{978-1-4503-9096-5/22/04}

\author{Yu Zheng$^{1*}$, Chen Gao$^{1\dagger}$, Jianxin Chang$^{2}$, Yanan Niu$^{2}$, Yang Song$^{2}$, Depeng Jin$^{1}$, Yong Li$^{1}$}
\thanks{$^*$Work done when interning at Kuaishou.}
\thanks{$\dagger$Chen Gao is the corresponding author (chgao96@gmail.com).}
\affiliation{%
 \institution{$^1$Beijing National Research Center for Information Science and Technology,}
 \institution{Department of Electronic Engineering, Tsinghua University}
 \institution{$^2$Beijing Kuaishou Technology Co., Ltd.}
 \country{}
}
\renewcommand{\authors}{Yu Zheng, Chen Gao, Jianxin Chang, Yanan Niu, Yang Song, Depeng Jin, Yong Li}

\renewcommand{\shortauthors}{Zheng, et al.}

\begin{document}

\title[Disentangling Long and Short-Term Interests for Recommendation]{Disentangling Long and Short-Term Interests\\ for Recommendation}

\begin{abstract}

Modeling user's long-term and short-term interests is crucial for accurate recommendation. 
However, since there is no manually annotated label for user interests, existing approaches always follow the paradigm of entangling these two aspects, which may lead to inferior recommendation accuracy and interpretability.
In this paper, to address it, we propose a \textbf{C}ontrastive learning framework to disentangle \textbf{L}ong and \textbf{S}hort-term interests for \textbf{R}ecommendation (\textbf{CLSR}) with self-supervision.
Specifically, we first propose two separate encoders to independently capture user interests of different time scales.
We then extract long-term and short-term \textit{interests proxies} from the interaction sequences, which serve as pseudo labels for user interests.
Then pairwise contrastive tasks are designed to supervise the similarity between interest representations and their corresponding interest proxies.
Finally, since the importance of long-term and short-term interests is dynamically changing, we propose to adaptively aggregate them through an attention-based network for prediction.
We conduct experiments on two large-scale real-world datasets for e-commerce and short-video recommendation. 
Empirical results show that our CLSR consistently outperforms all state-of-the-art models with significant improvements: GAUC is improved by over 0.01, and NDCG is improved by over 4\%. 
Further counterfactual evaluations demonstrate that stronger disentanglement of long and short-term interests is successfully achieved by CLSR.
The code and data are available at \url{https://github.com/tsinghua-fib-lab/CLSR}.

\end{abstract}



\maketitle

\input{1.intro.tex}

\input{2.method.tex}

\input{3.experiments.tex}

\input{4.related.tex}

\input{5.conclusion.tex}

\begin{acks}
This work is supported in part by The National Key Research and Development Program of China under grant 2020AAA0106000. This work is also supported in part by the National Natural Science Foundation of China under 61972223, U1936217, 61971267 and U20B2060.
\end{acks}

\balance
\bibliographystyle{ACM-Reference-Format}
\bibliography{sample-base}
\clearpage
\nobalance

\appendix
\input{6.appendix}

\end{document}

%% file: 1.intro.tex
\section{Introduction}\label{sec::intro}
With the deluge of information growing rapidly, recommender systems have been playing crucial roles in numerous online services, such as news \cite{an2019neural}, e-commerce \cite{zhou2018deep}, videos \cite{covington2016deep,li2019routing}, etc.
Specifically, recommender systems provide personalized contents by first inferring users' interests from their historical interactions and then retrieving items that meet these interests.
In practice, however, users' interest are difficult to track since they tend to have both stable long-term interests and dynamic short-term interests.
For example, a tech-savvy user may always be willing to browse electronics (long-term interest), while he may also exhibit interest in clothes in a short period (short-term interest).
As a result, accurately modeling and distinguishing users' long and short-term (\textbf{LS-term}) interests is critical.

Let us first review the literature.
Collaborative filtering (CF) based recommenders~\cite{koren2009matrix, rendle2012bpr, he2017neural, he2018outer, zhou2018deep} mainly capture the long-term interests and ignore the sequential features, thus they are limited in modeling the dynamic short-term interests.
Consequently, sequential models~\cite{hidasi2015session, zhu2017next, tang2018personalized, zhou2019deep} were proposed to exploit convolutional neural networks or recurrent neural networks to learn sequential features of user interests.
However, those methods tend to have short-term memory hence recommend items that are more relevant to users' recent behaviors.
As a result, recently, a series of approaches~\cite{zhao2018plastic, yu2019adaptive, an2019neural, lv2019sdm} were proposed to combine CF-based recommenders and sequential recommenders to cover both long-term and short-term interests.
Specifically, in these approaches, CF-based models such as matrix factorization are adopted for long-term interests, and sequential models are utilized to learn short-term interests.
However, whether LS-term interests can be effectively captured by the corresponding models is not guaranteed, since they impose no explicit supervision on the learned LS-term interests. 
In other words, the learned LS-term interests in those methods can be entangled with each other~\cite{locatello2019challenging}.

Overall speaking, disentangling LS-term interests faces the following challenges.
\begin{itemize}[leftmargin=*]

\item \textbf{First}, LS-term interests reflect quite different aspects of user preferences.
Specifically, long-term interests can be regarded as user's overall preferences which can remain stable for a long period of time, while short-term interests indicate a user's dynamic preferences that evolve rapidly according to recent interactions.
Therefore, learning a unified representation of LS-term interests is insufficient to capture such differences.
On the contrary, it is more proper to model the two aspects separately.
    \item \textbf{Second}, it is hard to obtain labeled data for learning LS-term interests. The collected behavior log data only always contains users' implicit feedback such as clicks.
Hence the separate modeling of LS-term interests lacks explicit supervision for distinguishing the two aspects.
    \item \textbf{Last}, for the final prediction of users' future interactions, both long and short-term interests should be taken into consideration.
Nevertheless, the importance of two kinds of interests varies on different user-item interactions.
For example, users' short-term interests are more important when they continuously browse similar items, while users' behaviors are largely driven by long-term interests when they switch to quite different items.
Therefore, it is critical but challenging to~\textit{adaptively} fuse these two aspects for predicting future interactions.

\end{itemize}

To address the above challenges, we propose a contrastive learning framework that disentangles LS-term interests leveraging the interaction sequences to build self-supervision signals.
Specifically, in order to independently capture LS-term interests, we propose to decompose each interaction into three mechanisms: long-term interests representation, short-term interests evolution, and interaction prediction.
We design two separate encoders with different dynamics over time to model LS-term interests respectively, which addresses the first challenge.
To overcome the key challenge of lacking labeled data for LS-term interests, we propose to use self-supervision \cite{chen2020simple}.
We first generate proxy representations for long/short-term interests by extracting users' entire historical interactions and recent interactions, respectively.
We then supervise the interest representations obtained from the two separate encoders to be more similar with their corresponding proxies than the opposite proxies, in a contrastive manner.
Different from existing methods which impose no explicit supervision on the learned LS-term interests \cite{yu2019adaptive, an2019neural}, our self-supervised approach can learn better-disentangled representations for LS-term interests and remove the dependency on labeled data.
With the disentangled interest representations, we design an attention-based fusion network that adaptively aggregates the two aspects for prediction, which solves the last challenge.

We evaluate the recommendation performance of our method on two real-world datasets.
Experimental results illustrate that CLSR outperforms state-of-the-art (SOTA) methods with significant improvements.
Specifically, AUC and GAUC are improved by over 0.02, and NDCG are improved by over 10.7\%, which can be considered as \textit{quite promising gain} by existing works \cite{song2019autoint,yu2019adaptive}.
To further investigate the effectiveness of the self-supervised disentanglement design, we conduct counterfactual evaluations with intervened historical interaction sequences which block long or short term interests.
The results demonstrate that CLSR achieves steadily stronger disentanglement of LS-term interests against SOTA methods.

In summary, the main contributions of this paper are as follows:
\begin{itemize}[leftmargin=*]
	\item We highlight the different dynamics of users' long and short-term interests, and take the pioneer step of disentangling the two factors is critical for accurate recommendation.

	\item We propose a contrastive learning framework to separately capture LS-term interests. Disentangled representations are learned with self-supervision by comparing with proxy representations constructed from the original interaction sequences. An attention-based fusion network is further designed which adaptively aggregates LS-term interests to predict interactions.
	
	\item We conduct extensive experiments on real-world datasets. 
	Experimental results validate that our proposed CLSR achieves significant improvements against SOTA methods.
	Further counterfactual analyses illustrate that much stronger disentanglement of LS-term interests can be achieved by CLSR.
	
\end{itemize}

The remainder of the paper is organized as follows.
We first formulate the problem in Section~\ref{sec::probdef} and introduce the proposed method in Section \ref{sec::method}.
We then conduct experiments in Section \ref{sec::experiments}, and review the related works in Section \ref{sec::related}.
Finally, we conclude the paper in Section \ref{sec::conclusion}.

%% file: 2.method.tex
\section{Problem Formulation}\label{sec::probdef}

\noindent\textbf{Notations.} Let $M$ denote the number of users, and $\{\bm{x^u}\}_{u=1}^{M}$ denote the interaction sequences for all users.
Each sequence $\bm{x^u} = [x^u_1, x^u_2, ... , x^u_{T_u}]$ denotes a list of items which are ordered by the corresponding interaction timestamps.
Here $T_u$ denotes the length of user $u$'s interaction history, and each item $x^u_t$ is in $[1, N]$, where $N$ denotes the number of items.

Since a user's interaction history $\bm{x^u}$ reflects both long and short-term interests, the recommender system will first learn LS-term interests from $\bm{x^u}$, and then predict future interactions based on the two aspects.
We then can formulate the problem of learning LS-term interests for recommendation as follows:

\noindent\textbf{Input:} The historical interaction sequences for all users $\{\bm{x^u}\}_{u=1}^{M}$.

\noindent\textbf{Output:} A predictive model that estimates the probability of whether a user will click an item, considering both LS-term interests.

\section{Methodology}\label{sec::method}
In this section, we elaborate on the proposed \textbf{C}ontrastive learning framework of \textbf{L}ong and \textbf{S}hort-term interests for \textbf{R}ecommendation (CLSR).

\subsection{User Interests Modeling}

\begin{figure}[t]
  \centering
  \includegraphics[width=\linewidth]{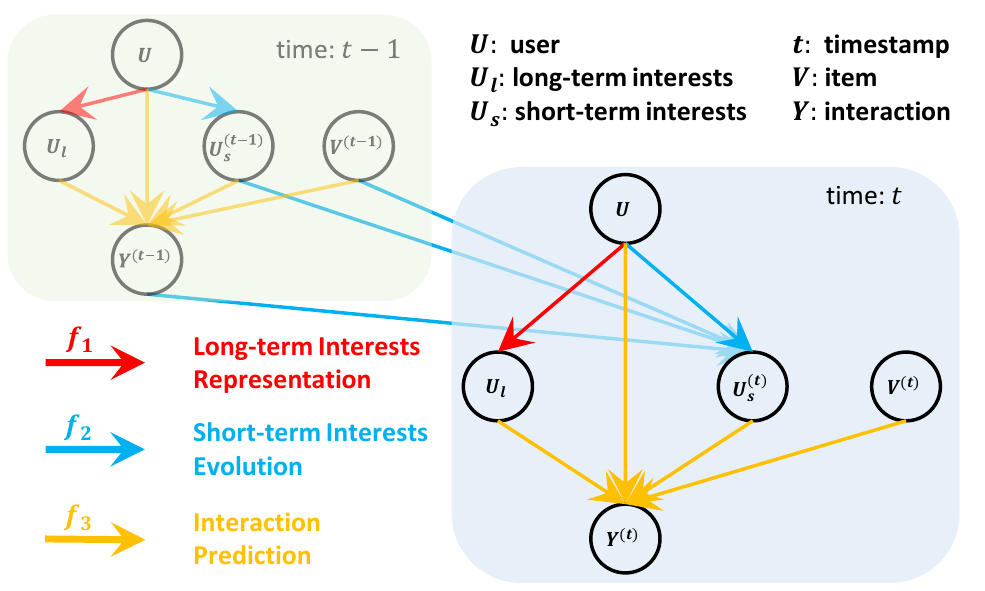}
  \vspace{-20px}
  \caption{User interests modeling $\zeta$ (best viewed in color) which consists of three mechanisms, namely long-term interests representation (red edges), short-term interests evolution (blue edges) and interaction prediction (yellow edges).}
  \label{fig::cg}
  \vspace{-15px}
\end{figure}

Since users' LS-term interests are quite different in terms of the dynamics over time, it is more appropriate to model the two aspects separately instead of using a unified representation to express them.
Specifically, long-term interests are relatively stable, while short-term interests are dynamic and changing frequently.
Meanwhile, each interaction is determined by both aspects as well as the target item.
Therefore, we propose to frame user interests modeling as the following three separate mechanisms:

\begin{numcases}{\zeta=}
U_l = f_1(U),  \\
U^{(t)}_s = f_2(U^{(t-1)}_s, V^{(t-1)}, Y^{(t-1)}, U),  \\
Y^{(t)} = f_3(U_l, U^{(t)}_s, V^{(t)}, U), 
\end{numcases}
where $f_1$, $f_2$ and $f_3$ are the underlying functions for user $U$'s long-term interests ($U_l$), short term interests ($U^{(t)}_s$) and interaction ($Y^{(t)}$) with item $V^{(t)}$. Current and last timestamps are denoted as $t$ and $t-1$, respectively.
It is worthwhile noting that $U$ denotes user profile, which contains the user ID and the interaction history $\bm{x^u}$.

The proposed user interests modeling $\zeta$ decomposes each interaction into three mechanisms: $f_1$ long-term interests representation, $f_2$ short-term interests evolution, and $f_3$ interaction prediction, which are briefly illustrated in Figure \ref{fig::cg}.
We now explain the details of the three mechanisms.

\begin{itemize}[leftmargin=*]
\item\textbf{Long-term Interests Representation in Eqn (1).} Long-term interests reflect a holistic view of user preferences, and hence it is stable and less affected by recent interactions.
In other words, long-term interests can be inferred from the entire historical interaction sequence, thus we include $U$ as the input of $f_1$, which contains the interaction history $\bm{x^u}$.

\item\textbf{Short-term Interests Evolution in Eqn (2).} Short-term interests are evolving as users continuously interact with recommended items~\cite{zhou2019deep}.
For example, users may establish new interests after clicking an item.
Meanwhile, users may also gradually lose certain interests.
That is to say, short-term interests are time-dependent variables, and thus in $f_2$, short-term interests $U^{(t)}_s$ at timestamp $t$ are evolved recursively from $U^{(t-1)}_s$, affected by the last interaction $Y^{t-1}$ with item $V^{(t-1)}$.

\item\textbf{Interaction Prediction in Eqn (3).} %
When predicting future interactions, whether long or short-term interests play a more important role depends on a wide variety of aspects, including the target item $V^{(t)}$ and the interaction history $\bm{x^u}$ of $U$ \cite{yu2019adaptive}.
Therefore, we fuse $U_l$ and $U^{(t)}_s$ according to $V^{(t)}$ and $U$ in an adaptive manner to accurately predict interactions.

\end{itemize}

Disentangling LS-term interests means that $U_l$ only captures long-term interests and $U_s$ models pure short-term interests.
Such disentanglement is helpful to achieve interpretable and controllable recommendation, since we can track and tune the importance of each aspect by adjusting the fusion weights.
Meanwhile, effective adjustment of LS-term interests requires the learned representations to only contain the information of the desired aspect.
Take the linear case as a toy example, suppose a recommendation model entangles LS-term interests as follows,
\begin{equation}
    U_l' = 0.6U_l + 0.4U_s, ~U_s' = 0.4U_l + 0.6U_s, 
\end{equation}
where $U_l'$ and $U_s'$ are the learned entangled interests.
Given the fusion weights (importance) of LS-term interests as 0.8 and 0.2 respectively, the actual fused interests are computed as follows,
\begin{equation}
    U_{fuse}' = 0.8U_l' + 0.2U_s' = 0.56U_l + 0.44U_s,
\end{equation}
which is quite different from the desired interests.

However, disentangling LS-term interests is challenging since there is no labeled data for $U_l$ and $U_s$.
We now elaborate on our contrastive learning framework which can achieve strong disentanglement with self-supervision.

\subsection{Our Self-supervised Implementation}

\begin{figure}[t]
  \centering
  \includegraphics[width=\linewidth]{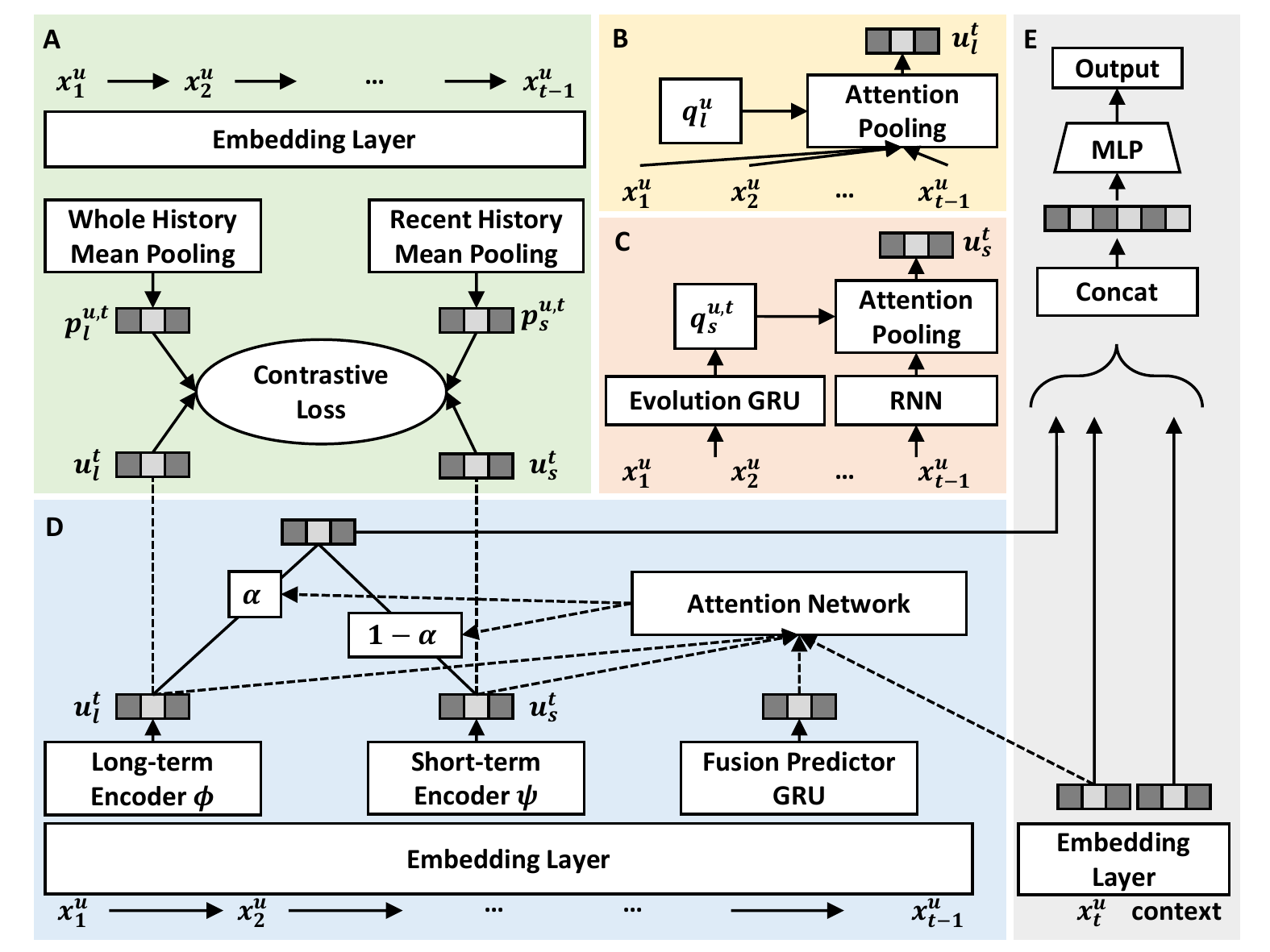}
  \vspace{-20px}
  \caption{Our proposed CLSR framework based on self-supervised learning. A) contrastive tasks on the similarity between representations and proxies of LS-term interests to enhance disentanglement; B) long-term interests encoder $\phi$; C) short-term interests encoder $\psi$; D) adaptive fusion of LS-term interests with attention on the target item and historical interactions; E) interaction prediction network.}
  \label{fig::pipeline}
  \vspace{-18px}
\end{figure}

In this section, we first provide two separate encoders to implement $f_1$ and $f_2$ which learn representations of LS-term interests.
Then we introduce our designed contrastive tasks to achieve disentanglement with self-supervision.
Last, we introduce the adaptive fusion model based on attention technique to accomplish $f_3$.
The overview of CLSR is illustrated in Figure \ref{fig::pipeline}.

\subsubsection{\textbf{Generating Query Vectors for LS-term Interests}}

Motivated by recent works \cite{zhao2018plastic, yu2019adaptive, an2019neural, lv2019sdm} that learn LS-term interests separately with two different models, we design two separate attentive encoders, $\phi$ and $\psi$, to capture the two aspects, respectively.
First, we generate query vectors for LS-term interests as follows,
\begin{align}
    &\bm{q^u_l} = \mathrm{Embed}(u), \label{eq::long}\\
    &\bm{q_s^{u,t}} = \mathrm{GRU}(\{x^u_1, \cdots, x^u_{t}\}),\label{eq::short}
\end{align}
where we use a look-up embedding table and a Gated Recurrent Unit (GRU) \cite{chung2014empirical} to capture different dynamics over time. 
In order to impose extra self-supervision on embedding similarity, all the embeddings need to be in the same semantic space.
Thus, we use the historical sequence of items as \textit{keys} of the attentive encoders, thus the obtained LS-term interests representations are in the same item embedding space as follows,
\begin{align}
	&\bm{u_l^t} = \phi(\bm{q^u_l}, \{x^u_1, \cdots , x^u_t\}), \\
	&\bm{u_s^t} = \psi(\bm{q_s^{u,t}}, \{x^u_1, \cdots , x^u_t\}),
\end{align}
where $\bm{u_l^t}$ and $\bm{u_s^t}$ are the learned representations of LS-term interests.
We now introduce the proposed encoders for LS-term interests.

\subsubsection{\textbf{Long-term Interests Encoder}}
Figure \ref{fig::pipeline} (B) illustrates the proposed long-term interests encoder $\phi$.
We use attention pooling to extract long-term interests representations, and the attention score of each item $x^u_j$ can be computed as follows,
\begin{align}
	&\bm{v}_j = \bm{W_l}\bm{E}(x^u_j), \\
	&\alpha_j = \tau_l(\bm{v_j}\|\bm{q_l^u}\|(\bm{v_j}-\bm{q_l^u})\|(\bm{v_j}\cdot \bm{q_l^u})), \label{eq::att1}\\
	&a_j = \frac{exp(\alpha_j)}{\sum_{i=1}^t{exp(\alpha_i)}},\label{eq::att2}
\end{align}
where $\bm{W_l}$ is a transformation matrix, $\tau_l$ is a multi-layer perceptrons (MLP) network, and $\|$ denotes the concatenation of embeddings.
The final learned long-term interests representation is a weighted aggregation of the entire interaction history, with weights computed from the above attentive network, formulated as follows,
\begin{equation}
	\bm{u_l^t} = \sum_{j=1}^t{a_j \cdot \bm{E}(x^u_j)}.
\end{equation}

\subsubsection{\textbf{Short-term Interests Encoder}}
Sequential patterns of user interaction play a crucial role in short-term interests modeling, thus we utilize another attentive network on top of a recurrent neural network (RNN).
Specifically, we feed the historical item embeddings to a RNN model and use the output of RNN as the keys, which can be formulated as follows,
\begin{align}
	&\{\bm{o^u_1}, ..., \bm{o^u_t}\} = \rho(\{\bm{E}(x^u_1), ... , \bm{E}(x^u_t)\}),\label{eq::rho} \\
	&\bm{v_j} = \bm{W_s}\bm{o^u_j},\end{align}
where $\bm{W_s}$ is a transformation matrix and $\rho$ represents a RNN model.
In Section \ref{sec::experiments}, we conduct experiments to evaluate different implementations of the RNN model, including LSTM \cite{hochreiter1997long}, GRU \cite{chung2014empirical} and Time4LSTM \cite{yu2019adaptive}.
Similar as Eqn (18) and (19), we use $\bm{q_s^{u,t}}$ as the query vector, and obtain attention scores $b_k$.
Then the learned representation for short-term interests can be computed as follows,
\begin{equation}
	\bm{u_s^t} = \sum_{j=1}^t{b_j \cdot \bm{o}^u_j}.
\end{equation}

Although separate encoders are adopted, disentanglement of LS-term interests is not guaranteed since $\bm{u_l^t}$ and $\bm{u_s^t}$ are extracted in an unsupervised manner \cite{locatello2019challenging}.
Particularly, there is no labeled data to supervise  the learned interests representations.
Therefore, we propose to design contrastive tasks which can achieve disentanglement with self-supervision and overcome the challenge of lacking labeled data.

\subsubsection{\textbf{Self-supervised Disentanglement of LS-Term Interests}.} %

As introduced previously, long-term interests provide a holistic view of user preferences which summarize the entire historical interactions, while short-term interests evolve dynamically over time which reflect recent interactions.
Therefore, we can obtain \textit{proxies} for LS-term interests from the interaction sequences themselves to supervise the two interests encoders.
Specifically, we calculate the mean representation of the entire interaction history as the proxy for long-term interests, and use the average representation of recent $k$ interactions as the proxy for short-term interests. 
Formally, the proxies of LS-term interests for a given user $u$ at timestamp $t$ can be calculated as follows,
\begin{align}
	&\bm{p_l^{u, t}} = \textbf{MEAN}(\{x^u_1, \cdots, x^u_t\}) = \frac{1}{t}\sum_{j=1}^{t}{\bm{E}(x^u_j)}, \\
	&\bm{p_s^{u, t}} = \textbf{MEAN}(\{x^u_{t-k+1}, \cdots , x^u_t\}) = \frac{1}{k}\sum_{j=1}^{k}{\bm{E}(x^u_{t-j+1})},
\end{align}
where $\bm{E}(x)$ means the embedding of item $x$.
Note that we only calculate proxies when the sequence length is longer than a threshold $l_t$, since there is no need to distinguish long and short-term if the whole sequence only contains a few items \cite{li2011logo}.
The threshold $l_t$, the length of the recent-behavior sequence $k$ are hyper-parameters in our method.
Furthermore, we use mean pooling here for its simplicity and the performance turns out to be good enough.
In fact, our self-supervised paradigm is capable of exploiting more complex design for proxies which we leave for future work.

With proxies serving as \textit{labels}, we can utilize them to supervise the disentanglement of LS-term interests.
Specifically, we perform contrastive learning between the encoder outputs and proxies, which requires the learned representations of LS-term interests to be more similar to their corresponding proxies than the opposite proxies.
We illustrate the contrastive tasks in Figure \ref{fig::pipeline} (A).
Formally, there are four contrastive tasks as follows,
\begin{align}
	&sim(\bm{u_l^t}, \bm{p_l^{u, t}}) > sim(\bm{u_l^t}, \bm{p_s^{u, t}}), \label{eq::con1}\\
	&sim(\bm{p_l^{u, t}}, \bm{u_l^t}) > sim(\bm{p_l^{u, t}}, \bm{u_s^t}), \label{eq::con2}\\
	&sim(\bm{u_s^t}, \bm{p_s^{u, t}}) > sim(\bm{u_s^t}, \bm{p_l^{u, t}}), \label{eq::con3}\\
	&sim(\bm{p_s^{u, t}}, \bm{u_s^t}) > sim(\bm{p_s^{u, t}}, \bm{u_l^t}), \label{eq::con4}
\end{align}
where Eqn (\ref{eq::con1})-(\ref{eq::con2}) supervise long-term interests, and Eqn (\ref{eq::con3})-(\ref{eq::con4}) supervise short-term interests, and $sim(\cdot,\cdot)$ measures embedding similarity.
Take long-term interests modeling as an example, Eqn (\ref{eq::con1}) encourages the learned long-term interests representation, $\bm{u_l^t}$, to be more similar to the long-term proxy, $\bm{p_l^{u, t}}$, than to the short-term proxy, $\bm{p_s^{u, t}}$.
Meanwhile, Eqn (\ref{eq::con2}) requires that $\bm{u_l^t}$ is closer to $\bm{p_l^{u, t}}$ compared with the short-term interests representation, $\bm{u_s^t}$. 
With four symmetric contrastive tasks on the similarity between encoder outputs and proxies, we add self-supervision on LS-term interests modeling which can achieve stronger disentanglement compared with existing unsupervised approaches.

We implement two pairwise loss functions based on Bayesian Personalized Ranking (BPR) \cite{rendle2012bpr} and triplet loss to accomplish contrastive learning in Eqn (\ref{eq::con1})-(\ref{eq::con4}).
Formally, the two loss functions, which use inner product and Euclidean distance to capture embedding similarity, are computed as follows,
\begin{align}
	&\mathcal{L}_{\text{bpr}}(a, p, q) = \sigma(\langle a, q\rangle - \langle a, p\rangle), \\
	&\mathcal{L}_{\text{tri}}(a, p, q) = \text{max}\{d(a, p) - d(a, q) + m, 0\},
\end{align}
where $\sigma$ is the \textit{softplus} activation function, $\langle\cdot, \cdot\rangle$ denotes inner product of two embeddings, $d$ denotes the Euclidean distance, and $m$ denotes a positive margin value.
Both $\mathcal{L}_{bpr}$ and $\mathcal{L}_{tri}$ are designed for making the anchor $a$ more similar to the positive sample $p$ than the negative sample $q$.
Thus the contrastive loss for self-supervised disentanglement of LS-term interests can computed as follows,
\begin{equation}
    \mathcal{L}_{\text{con}}^{u, t} = f(\bm{u}_l, \bm{p}_l, \bm{p}_s) + f(\bm{p}_l, \bm{u}_l, \bm{u}_s) + f(\bm{u}_s, \bm{p}_s, \bm{p}_l) + f(\bm{p}_s, \bm{u}_s, \bm{u}_l)\label{eq::L_con}
\end{equation}
where we omit the superscript of interest representations and proxies, and $f$ can be either $\mathcal{L}_{bpr}$ or $\mathcal{L}_{tri}$.

\noindent\textbf{Remark.}
Users' LS-term interests can also overlap with each other to some extent.
For example, a user who only purchases clothes on an e-commerce application tends to have consistent LS-term interests.
Therefore, unlike existing disentangled recommendation approaches \cite{zheng2021disentangling,wang2020disentangled} which add an independence constraint forcing the learned disentangled factors to be dissimilar with each other, we do not include such regularization term and only supervise the learned representations of LS-term interests to be similar with their corresponding proxies.
This is also why we do not use loss functions like InfoNCE \cite{oord2018representation} which impose too strong punishment on the similarity between opposite encoders and proxies.

In summary, we implement two separate encoders $\phi$ and $\psi$ to learn representations for LS-term interests, respectively.
In order to achieve disentanglement of LS-term interests, we compute proxies from the historical interaction sequences.
We further propose contrastive-learning loss functions that guide the two encoders only to capture the desired aspect in a self-supervised manner.

\subsubsection{\textbf{Adaptive Fusion for Interaction Prediction}}

With the learned disentangled representations by self-supervised learning, how to aggregate the two aspects to predict interactions remains a challenge.
Simple aggregators, such as sum and concatenation, assume that contributions of LS-term interests are fixed, which is invalid in many cases.
In fact, whether long or short-term one is more important depends on the historical sequence.
For example, users are mainly driven by short-term interests when they are continuously browsing items from the same category.
Meanwhile, the importance of LS-term interests also depends on the target item.
For instance, a sports lover may still click on a recommended bicycle due to long-term interests, even after he/she browses several books.
Therefore, we include both the historical sequence and the target item as input of the aggregator, where historical sequence is compressed with a GRU.
The proposed attention-based adaptive fusion model is illustrated in Figure \ref{fig::pipeline} (D), which dynamically determines the importance of LS-term interests to aggregate  $\bm{u_l^t}$ and $\bm{u_s^t}$.
Formally, the final fused interests are obtained as follows,
\begin{align}
	&\bm{h^u_t} = \mathrm{GRU}(\{\bm{E}(x^u_1), ... , \bm{E}(x^u_t)\}), \\
	&\alpha = \sigma(\tau_f(\bm{h}^u_t\|\bm{E}(x^u_{t+1})\|\bm{u_l^t}\|\bm{u_s^t}), \\
	&\bm{u}^t = \alpha\cdot \bm{u_l^t} + (1 - \alpha)\cdot \bm{u_s^t},
\end{align}
where $\sigma$ is the \textit{sigmoid} activation function, and $\tau_f$ is a MLP for fusion.
Here $\alpha$ denotes the estimated fusion weight based on historical interactions, target item, and user's LS-term interests.

To predict the interaction, we use the widely adopted two-layer MLP \cite{yu2019adaptive} shown in Figure \ref{fig::pipeline} (E).
Then the estimated score given a user $u$ and an item $v$ at timestamp $t+1$ can be predicted as follows,
\begin{equation}
	\hat{y}^{t+1}_{u,v} = \mathrm{MLP}(\bm{u}^t\|\bm{E}(v)).
\end{equation}

Following the existing works' settings\cite{yu2019adaptive}, we use the negative log-likelihood loss function as follows,
\begin{equation}
	\mathcal{L}_{\text{rec}}^{u, t} = - \frac{1}{N}\sum_{v \in \mathcal{O}}{y^{t+1}_{u,v}\log(\hat{y}^{t+1}_{u,v}) + (1 - y^{t+1}_{u,v})\log(1 - \hat{y}^{t+1}_{u,v})},
\end{equation}
where $\mathcal{O}$ is the set composed of training pairs of one positive item $x^u_{t+1}$ and $N-1$ sampled negative items.
We train the model in an end-to-end manner with multi-task learning on two objectives. Specifically, the joint loss function with a hyper-parameter $\beta$ to balance objectives, can be formulated as follows,
\begin{equation}
	\mathcal{L} =  \sum_{u=1}^M\sum_{t=1}^{T_u}{(\mathcal{L}_{\text{rec}}^{u, t} + \beta \mathcal{L}_{\text{con}}^{u, t})} + \lambda \|\Theta\|_2,
\end{equation}
where $\lambda \|\Theta\|_2$ denotes the $L2$ regularization for addressing over-fitting. The computation complexity of our implementation is $\mathcal{O}((M+N)d + Q)$ where $Q$ denotes the complexity of MLP and GRU, which is on par with the state-of-the-art SLi-Rec method \cite{yu2019adaptive}.

%% file: 3.experiments.tex
\section{Experiments}\label{sec::experiments}

\begin{table}
	\caption{Statistics of the two datasets used in experiments.}
	\vspace{-10px}
	\label{tab::dataset}
	\begin{tabular}{cccccc}
    \toprule
    Dataset & Users & Items & Instances & Average Length \\ 
    \midrule
    Taobao & 36,915 & 64,138 & 1,471,155 & 39.85 \\
    Kuaishou & 60,813 & 292,286 & 14,952,659 & 245.88 \\
    \bottomrule
    \end{tabular}
    \vspace{-10px}
\end{table}

In this section, we conduct experiments to show the effectiveness of the proposed contrastive learning framework. 
Specifically, we aim to answer the following research questions,
\begin{itemize}[leftmargin=*]
	\item \textbf{RQ1:} How does the proposed framework perform compared with state-of-the-art recommendation models?
	\item \textbf{RQ2:} Can CLSR achieves stronger disentanglement of LS-term interests against existing unsupervised baselines?
	\item \textbf{RQ3:} What is the effect of different components in CLSR?
\end{itemize}

\begin{table*}[t!]
    \caption{Overall performance on Taobao and Kuaishou datasets. \underline{Underline} means the best two baselines, \textbf{bold} means \textit{p}-value < 0.05, * means \textit{p}-value < 0.01, and ** means \textit{p}-value < 0.001.}
    \vspace{-10px}
    \label{tab::overall}
    \begin{tabular}{cc|cccc|cccc}
      \toprule
      \multicolumn{2}{c|}{Dataset} & \multicolumn{4}{c|}{Taobao} & \multicolumn{4}{c}{Kuaishou} \\
      \hline
      Category & Method & AUC & GAUC & MRR & NDCG@2 & AUC & GAUC & MRR & NDCG@2\\
      \midrule
      \multicolumn{1}{c}{\multirow {3}{*}{Long-term}} & NCF & 0.7128 & 0.7221 & 0.1446 & 0.0829 & 0.5559 & 0.5531 & 0.7734 & 0.8327 \\
      & DIN & 0.7637 & 0.8524 & 0.3091 & 0.2352 & 0.6160 & 0.7483 & 0.8863 & 0.9160  \\
      & LightGCN & 0.7483 & 0.7513 & 0.1669 & 0.1012 & 0.6403 & 0.6407 & 0.8175 & 0.8653 \\
      \hline
      \multicolumn{1}{c}{\multirow {5}{*}{Short-term}} & Caser & 0.8312 & 0.8499 & 0.3508 & 0.2890 & 0.7795 & 0.8097 & 0.9100 & 0.9336 \\
      & GRU4REC & 0.8635 & 0.8680 & 0.3993 & \underline{0.3422} & 0.8156 & \underline{0.8298} & \underline{0.9166} & \underline{0.9384} \\
      & DIEN & 0.8477 & \underline{0.8745} & \underline{0.4011} & 0.3404 & 0.7037 & 0.7800 & 0.9030 & 0.9284 \\
      & SASRec & 0.8598 & 0.8635 & 0.3915 & 0.3340 & \underline{0.8199} & 0.8293 & 0.9161 & 0.9380 \\
      & SURGE & \underline{0.8906} & \underline{0.8888} & \underline{0.4228} & \underline{0.3625} & \underline{0.8525} & \underline{0.8610} & \underline{0.9316} & \underline{0.9495} \\
      \hline
      \multicolumn{1}{c}{\multirow {2}{*}{LS-term}} & SLi-Rec & \underline{0.8664} & 0.8669 & 0.3617 & 0.2971 & 0.7978 & 0.8128 & 0.9075 & 0.9318 \\
      & Ours & $\textbf{0.8953}^{**}$ & $\textbf{0.8936}^{**}$ & $\textbf{0.4372}^{**}$ & $\textbf{0.3788}^{**}$ & \textbf{0.8563} & \textbf{0.8718} & $\textbf{0.9382}^*$ & $\textbf{0.9544}^*$ \\
      \bottomrule
    \end{tabular}
    \vspace{-10px}
\end{table*}

\noindent\textbf{Datasets.}
We conduct experiments on two datasets collected from real-world e-commerce and video platforms, Taobao\footnote{\url{https://www.taobao.com}} and Kuaishou\footnote{\url{https://www.kuaishou.com}}.
Basic statistics of the two datasets are summarized in Table \ref{tab::dataset}, where Average Length indicates the average length of user interaction sequences.
We leave the details of the datasets in Section \ref{app::dataset}.

\noindent{\textbf{Baselines and Metrics.}} We compare CLSR with state-of-the-art methods.
With respect to long-term interests modeling, we include \textbf{NCF} \cite{he2017neural}, \textbf{DIN} \cite{zhou2018deep} and \textbf{LightGCN}\cite{he2020lightgcn}.
For short-term interests modeling, we compare with \textbf{Caser} \cite{tang2018personalized}, \textbf{GRU4REC} \cite{hidasi2015session}, \textbf{DIEN} \cite{zhou2019deep}, \textbf{SASRec} \cite{kang2018self} and \textbf{SURGE} \cite{chang2021sequential}.
We also include \textbf{SLi-Rec} \cite{yu2019adaptive} which is the state-of-the-art model of LS-term interests modeling.
We evaluate the models with two widely-adopted accuracy metrics including \textbf{AUC} and \textbf{GAUC} \cite{zhou2018deep}, as well as two commonly used ranking metrics \textbf{MRR} and \textbf{NDCG@K}. 
We leave the details of baselines, implementations, and hyper-parameters in Section \ref{app::baseline}-\ref{app::implementation}.

\subsection{Overall Performance Comparison (RQ1)}

We illustrate the overall performance on two adopted datasets in Table \ref{tab::overall}.
From the results, we have the following observations:
\begin{itemize}[leftmargin=*]
    \item \textbf{Short-term models generally performs better than long-term models.} %
    Long-term models fail to capture temporal patterns of user interactions, hence their performance is rather poor. 
    From the results, we can observe that AUC of NCF, DIN, and LightGCN are all less than 0.8 on Taobao dataset and less than 0.7 on Kuaishou dataset. 
    On the other hand, short-term models outperform long-term models in most cases.
    For example, SURGE is the best baseline on both datasets, which uses graph convolutional propagation and graph pooling to capture the dynamics of user interests.
    The better performance of short-term models comes from their ability to capture the sequential pattern of user interactions. 
    In fact, we conduct data analysis on the interaction sequences and discover that, in average, over 31\% of interacted items are of the same category as the previous item, which verifies the sequential pattern and explains the better performance of short-term models.
    \item \textbf{Joint modeling of LS-term interests does not always bring performance gains.} 
    SLi-Rec is the SOTA approach that models both LS-term interests. 
    However, the two aspects are entangled with each other, which increases model redundancy and leads to inferior accuracy. 
    Results demonstrate that SLi-Rec is not consistently effective across different metrics and datasets. 
    For example, SLi-Rec is the best baseline on Taobao dataset with respect to AUC, but its ranking performance is poorer than GRU4REC by about 10\%, indicating that it is insufficient to disentangle LS-term interests with no explicit supervision.
    \item \textbf{Disentangled modeling of LS-term interests can achieve significant improvements.} 
 	CLSR outperforms baselines with significant progress.
	Specifically, CLSR improves GAUC by about 0.005 (\textit{p}-value < 0.001) on Taobao dataset and 0.01 (\textit{p}-value < 0.05) on Kuaishou dataset, agaisnt SOTA methods. 
	Besides, NDCG is improved by about 5\% on Taobao dataset.
	The consistent and significant progress indicate that disentangling LS-term interests is critical for accurate recommendation.

\end{itemize}

\subsection{Study on Disentanglement of Long and Short-Term Interests (RQ2)}
Both SLi-Rec and CLSR explicitly model LS-term interests, however, CLSR achieves the best performance while SLi-Rec shows inferior accuracy.
We argue that it is because SLi-Rec entangles LS-term interests which increases the internal dependency of the model and leads to poor performance.
On the contrary, CLSR disentangles LS-term interests with the help of self-supervision.
In this section, we empirically prove that stronger disentanglement of LS-term interests is indeed achieved by CLSR.

\subsubsection{\textbf{Performance of One-side Interests}}

\begin{figure}[t!]
\centering
\includegraphics[width=\linewidth]{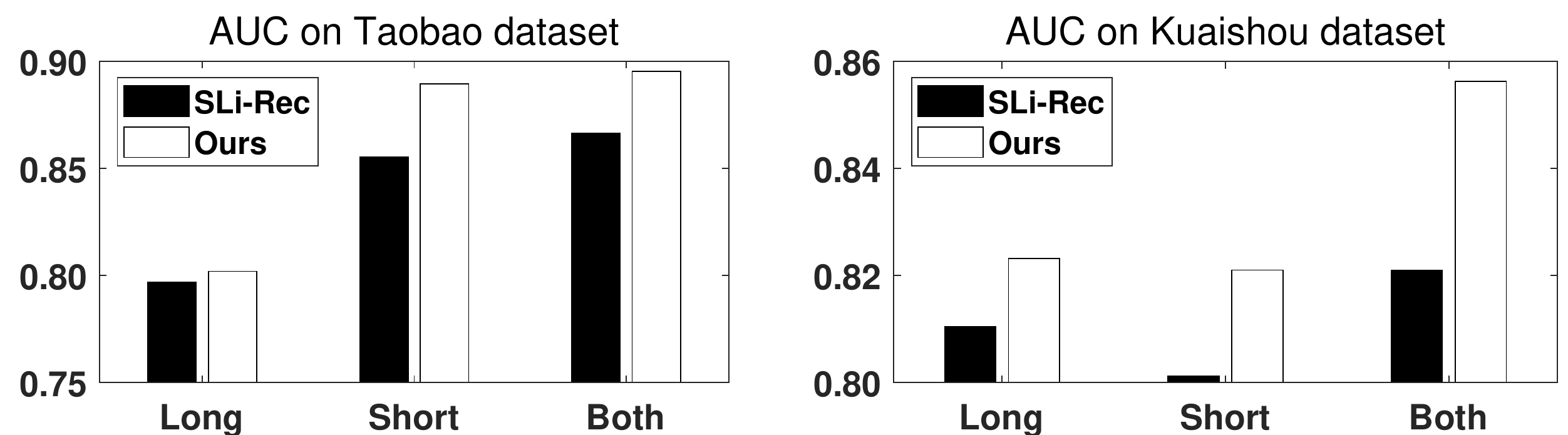}
\vspace{-20px}
\caption{Comparison of using single and both interests between CLSR and Sli-Rec.}
\label{fig::long_short_both}
\vspace{-10px}
\end{figure}

In CLSR, we utilize two separate representations for LS-term interests.
Therefore, it is crucial that each side only captures the desired single aspect.
In order to evaluate the effectiveness of each side, we reserve one-side interests and discard the other side of CLSR and SLi-Rec.
Results on two datasets are illustrated in Figure \ref{fig::long_short_both}, from which we can observe that CLSR outperforms SLi-Rec in all cases.
Specifically, on Taobao dataset, CLSR improves AUC against SLi-Rec by about 0.03 with short-term interests and full interests.
On Kuaishou dataset, the improvements of AUC are about 0.1, 0.2, and 0.4 for long-term interests, short-term interests, and full interests, respectively.
It indicates that CLSR attains more meaningful representations for both LS-term interests.
Moreover, for both methods on both datasets, combining LS-term interests achieves better performance than using one-side interests.
This further supports our motivation to model both long and short-term interests for accurate recommendation.

\subsubsection{\textbf{Counterfactual Evaluation}}

Learning disentangled representations of underlying factors is very helpful especially when the importance of different factors changes \cite{zheng2021disentangling,scholkopf2019causality,scholkopf2021toward}.
For example, behaviors of higher costs, such as purchase (cost of money) in Taobao dataset and like (cost of time) in Kuaishou dataset, tend to be more driven by users' long-term interests, 
and behaviors of lower costs such as click in both datasets indicate more about short-term interests, which has been acknowledged by existing works \cite{grbovic2018real}.
Therefore, to investigate whether CLSR achieves disentanglement of LS-term interests, we design counterfactual evaluations where the importance of different interests changes.
Specifically, we use models well-trained on click data to predict both clicked items and purchased/liked items, where the importance of LS-term interests is different.
Since purchase/like behavior reflects more long-term interests, the importance of long-term interests is supposed to be higher when predicting purchase/like than click.
In other words, when the model predicts purchase/like behavior, it is expected that the attention weight for long-term interests when fusing the two aspects, \textit{i.e.} $\alpha$, to be also larger than predicting click behavior.

Table \ref{tab::counter} illustrates the AUC and the average of $\alpha$ for clicked items and purchased/liked items.
We have the following findings:
\begin{itemize}[leftmargin=*]
	\item CLSR outperforms SLi-Rec for all behaviors. Although predicting purchase/like with models trained on click data is challenging, AUC of CLSR is significantly larger than SLi-Rec by over 0.03.
	Meanwhile, the average $\alpha$ of CLSR is much lower in all cases, unlike SLi-Rec whose average $\alpha$ is even over 0.7 on Kuaishou dataset.
	In fact, low $\alpha$ in CLSR is consistent with previous findings in Table \ref{tab::overall} that long-term interests are less important than short-term interests, which means that LS-term interests are successfully disentangled in CLSR.
	On the contrary, high $\alpha$ in SLi-Rec indicates that the learned long-term interests representations contain much information of the undesired short-term interests, \textit{i.e.} the two aspects entangles with each other.
	\item Since purchase/like reflects more long-term interests than click, $\alpha$ is supposed to be also larger when predicting purchase/click. 
	On Taobao dataset, $\alpha$ of CLSR for purchase behavior is larger than click by about 4\%.
	However, for SLi-Rec, $\alpha$ for purchase is even less than click by over 6\%. 
	On Kuaishou dataset, though $\alpha$ for like is larger than click in both SLi-Rec and CLSR, the relative increment of $\alpha$ for CLSR is over two times larger than SLi-Rec (+9.06\% v.s. +3.91\%). 
	This further validates that CLSR achieves much stronger disentanglement of LS-term interests.
\end{itemize}

\begin{table}[t!]
\small
    \caption{Comparison between CLSR and SLi-Rec on predicting click and purchase/like.}
    \vspace{-10px}
    \label{tab::counter}
    \begin{tabular}{c|c|cc|cc}
      \toprule
      \multicolumn{1}{c|}{\multirow {2}{*}{Dataset}} & \multicolumn{1}{c|}{\multirow {2}{*}{Method}} & \multicolumn{2}{c|}{Click} & \multicolumn{2}{c}{Purchase/Like} \\
      \multicolumn{1}{c|}{} & \multicolumn{1}{c|}{} & AUC & AVG($\alpha$) & AUC & AVG($\alpha$) \\
      \midrule
      \multicolumn{1}{c|}{\multirow {2}{*}{Taobao}} & SLi-Rec & 0.8572 & 0.4651 & 0.8288 & 0.4350 (-6.47\%) \\
      \multicolumn{1}{c|}{} & CLSR & 0.8885 & 0.3439 & 0.8616 & 0.3568 (\bf{+3.75\%}) \\
      \hline
      \multicolumn{1}{c|}{\multirow {2}{*}{Kuaishou}} & SLi-Rec & 0.8153 & 0.7259 & 0.7924 & 0.7543 (+3.91\%) \\
      \multicolumn{1}{c|}{} & CLSR & 0.8618 & 0.2528 & 0.7946 & 0.2757 (\bf{+9.06\%}) \\
      \bottomrule
    \end{tabular}
    \vspace{-10px}
\end{table}
\begin{table}[t!]
    \caption{Counterfactual evaluation under shuffle protocol.}
    \vspace{-10px}
    \label{tab::counter_shuffle}
    \begin{tabular}{c|c|cc|cc}
      \toprule
      \multicolumn{1}{c|}{\multirow {2}{*}{Dataset}} & \multicolumn{1}{c|}{\multirow {2}{*}{Method}} & \multicolumn{2}{c|}{Click} & \multicolumn{2}{c}{Purchase/Like} \\
      \multicolumn{1}{c|}{} & \multicolumn{1}{c|}{} & AUC & MRR & AUC & MRR  \\
      \midrule
      \multicolumn{1}{c|}{\multirow {2}{*}{Taobao}} & SLi-Rec & 0.8092 & 0.2292 & 0.8480 & 0.3151 \\
      \multicolumn{1}{c|}{} & CLSR & \bf{0.8413} & \bf{0.2744} & \bf{0.8790} & \bf{0.4194} \\
      \hline
      \multicolumn{1}{c|}{\multirow {2}{*}{Kuaishou}} & SLi-Rec & 0.7992 & 0.9088 & 0.8165 & 0.9113 \\
      \multicolumn{1}{c|}{} & CLSR & \bf{0.8431} & \bf{0.9380} & \bf{0.8197} & \bf{0.9167} \\
      \bottomrule
    \end{tabular}
    \vspace{-10px}
\end{table}

\begin{figure}[t!]
  \centering
  \includegraphics[width=\linewidth]{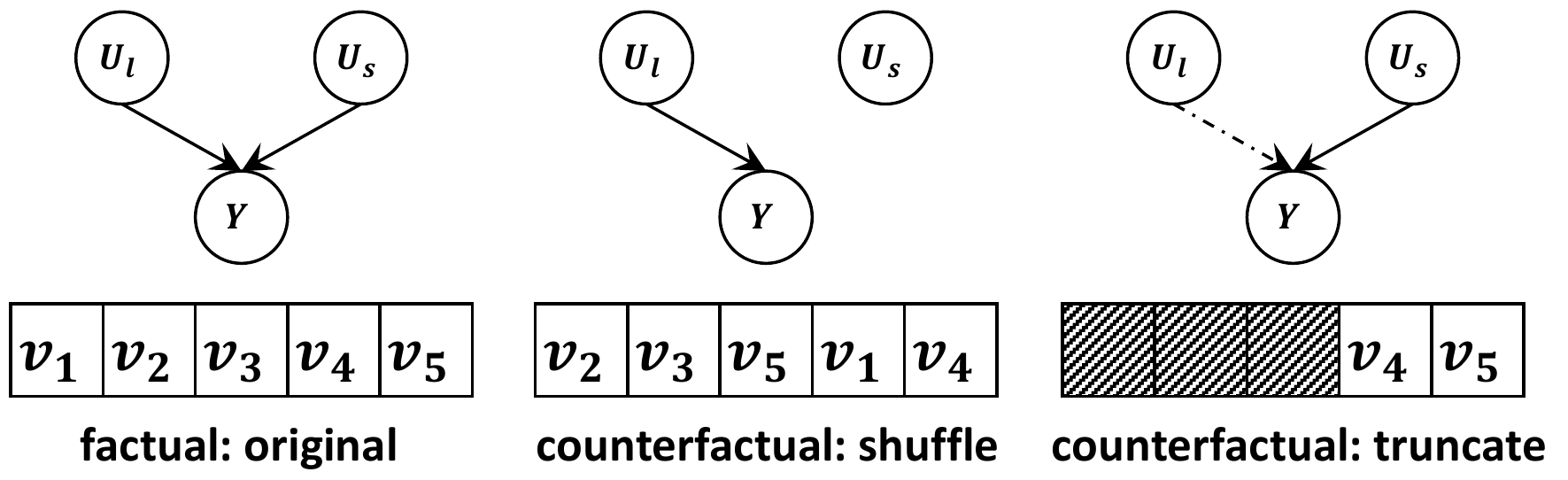}
  \vspace{-20px}
  \caption{Counterfactual evaluation. Shuffle: short-term interests are removed by shuffling. Truncate: long-term interests are weakened by discarding early history.}
  \label{fig::counter}
  \vspace{-15px}
\end{figure}

Meanwhile, we also evaluate under special cases where long or short-term interests are blocked by re-arranging interaction sequences with two protocols, namely shuffle and truncate, as illustrated in Figure \ref{fig::counter}.
The details are as follows.
\begin{itemize}[leftmargin=*]
	\item \textbf{Shuffle}: The historical sequence is randomly shuffled, and thus short-term interests are removed under this protocol.
	\item \textbf{Truncate}: Early history is discarded and only recent history is available. Thus long-term interests are weakened.
\end{itemize}

Table \ref{tab::counter_shuffle} shows the results under shuffle protocol on two datasets.
Since shuffling operation blocks short-term interests, predicting click behavior is much more difficult than the original case, while predicting purchase behavior is relatively easier.
We can observe that the results in Table \ref{tab::counter_shuffle} compared with Table \ref{tab::counter} is consistent with the expectation.
Specifically, for both SLi-Rec and CLSR, AUC decreases by over 0.04 and increases by about 0.02 on click-prediction task and purchase/like-prediction task, respectively.
Meanwhile, CLSR improves the AUC of click-prediction by over 0.04, and improves the MRR of purchase-prediction by over 30\%, against SLi-Rec.
Although short-term interests are invalid under this protocol, CLSR can still achieves better performance since LS-term interests are disentangled and long-term interests can still take effect.

\begin{figure}[t!]
  \centering
  \includegraphics[width=\linewidth]{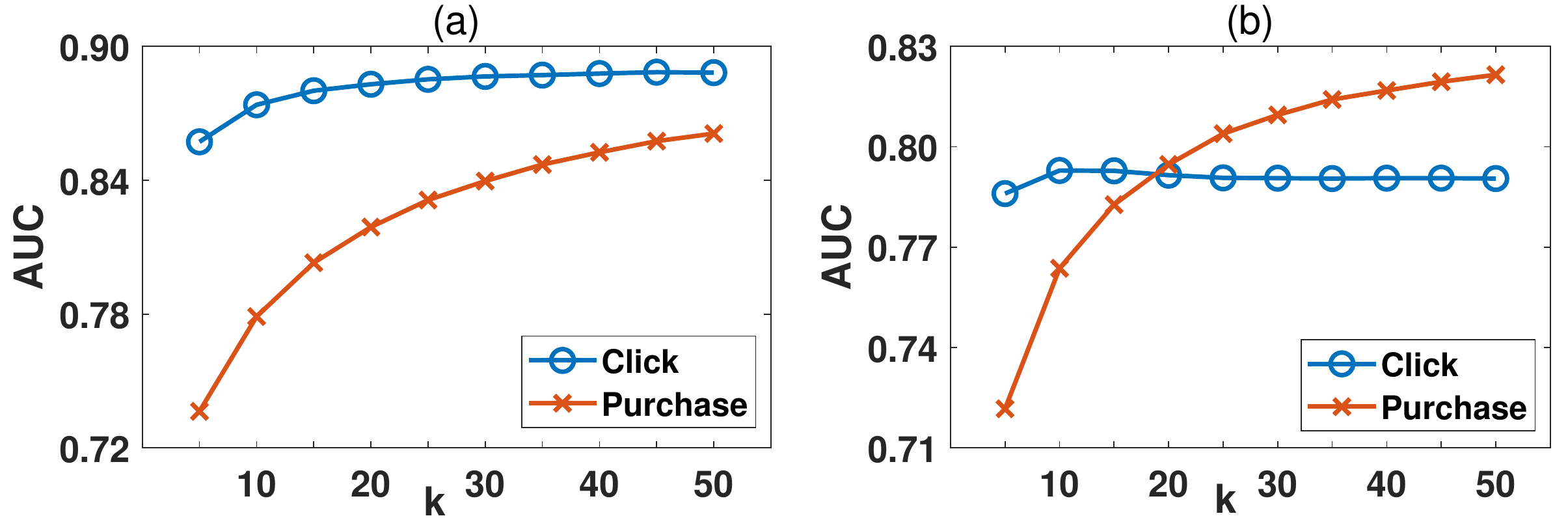}
  \vspace{-20px}
  \caption{Counterfactual evaluation under truncate protocol. (a) CLSR. (b) CLSR with only long-term interests.}
  \label{fig::truncate}
   \vspace{-15px}
\end{figure}

We further present the results of CLSR under truncate protocol with varying available length ($k$) of historical sequences in Figure \ref{fig::truncate} (a).
We can observe that the performance of purchase prediction improves significantly as $k$ grows.
Meanwhile, the performance of click prediction increases much slower when $k$ grows larger.
This observation verifies our assumption that short-term interests can be effectively captured by mining the recent history, while for long-term interests, it is essential to take the entire history into consideration.
In addition, we also show the performance of only using long-term interests representation under truncate protocol in Figure \ref{fig::truncate} (b).
We can find that the accuracy of purchase-prediction increases drastically as $k$ getting larger, while the accuracy of click-prediction is barely changed.
The different trends of click and purchase tasks confirm that the learned long-term interests representations only capture the desired interests and distill short-term interests.


In summary, by comparing CLSR and SLi-Rec, which both explicitly model LS-term interests, we empirically show that disentanglement of the two aspects is the reason of better recommendation performance.
Moreover, it is insufficient to disentangle LS-term interests in an unsupervised way, and CLSR effectively overcomes the challenge of lacking labeled data with self-supervision.

\subsection{Ablation and Hyper-parameter Study (RQ3)}

\subsubsection{\textbf{Contrastive Learning}}
\begin{figure}[t]
  \centering
  \includegraphics[width=\linewidth]{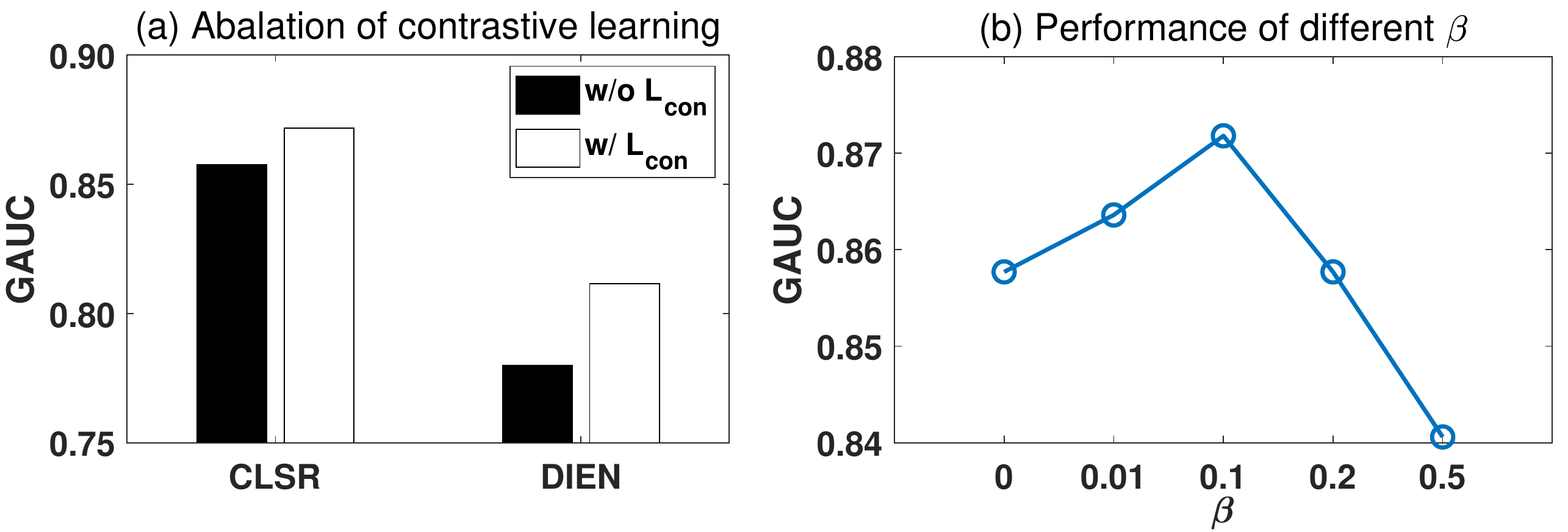}
  \vspace{-25px}
  \caption{(a) Ablation study of contrastive loss. (b) Hyper-parameter study of $\beta$.}
  \label{fig::contrastive}
  \vspace{-15px}
\end{figure}

Contrastive tasks on the similarity between learned representations and proxies for LS-term interests help achieve stronger disentanglement than existing unsupervised methods.
We conduct ablation study to compare the performance of CLSR with and without the contrastive loss $\mathcal{L}_{\text{con}}$.
In addition, we also evaluate the performance of replacing the short-term interests encoder $\psi$ with DIEN.
Figure \ref{fig::contrastive} (a) illustrates the results on Kuaishou dataset.
We can find that GAUC of CLSR drops over 0.01 after removing the contrastive tasks which verifies the necessity of self-supervision.
Meanwhile, adding self-supervision can also significantly improve the performance of DIEN, which means that CLSR can serve as a general framework to disentangle LS-term interests for existing recommendation models.
We also investigate the performance under different loss weights of $\mathcal{L}_{\text{con}}$.
Figure \ref{fig::contrastive} (b) illustrates the results on Kuaishou dataset.
We can observe that 0.1 is an optimal value, and too large $\beta$ may contradict with the main interaction prediction task which leads to low accuracy.

\begin{figure}[t]
  \centering
  \includegraphics[width=\linewidth]{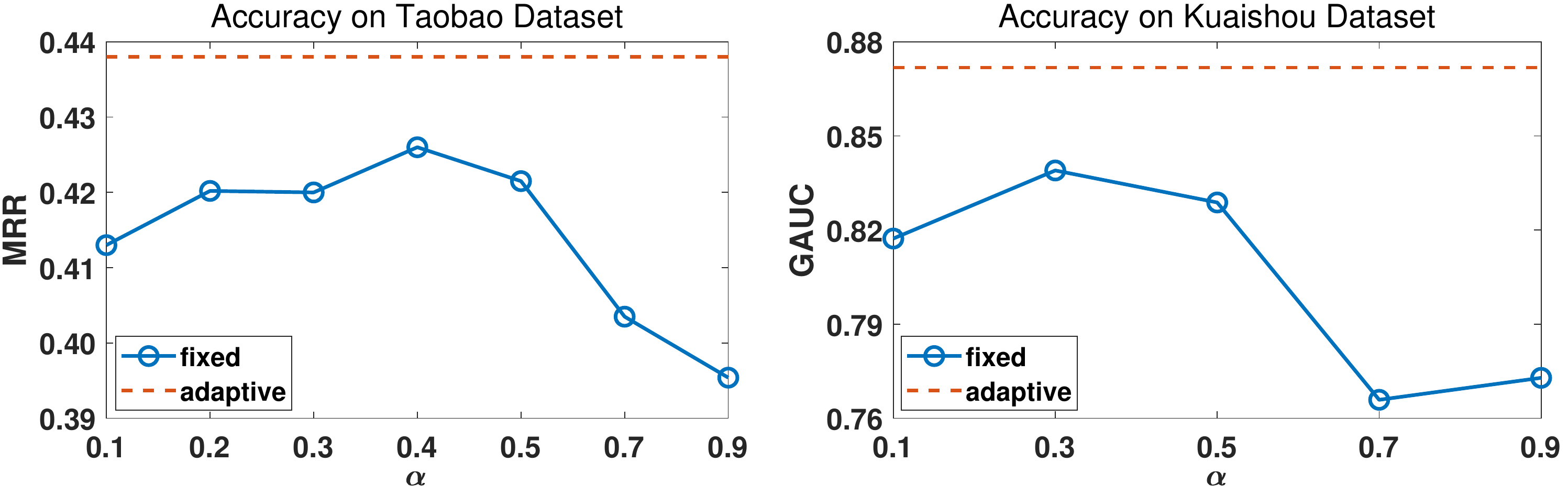}
  \vspace{-25px}
  \caption{Comparison between adaptive and fixed fusion.}
  \label{fig::manual_alpha}
  \vspace{-15px}
\end{figure}

\subsubsection{\textbf{Adaptive Fusion of LS-Term Interests}}
In CLSR, we propose to aggregate LS-term interests adaptively according to the target item and the historical sequence.
Here we investigate whether this adaptive fusion is effective.
To be specific, we compare with a static version, which means using a fixed $\alpha$ when combining the two aspects.
Figure \ref{fig::manual_alpha} shows the recommendation performance on two datasets, where the dashed line represents the performance of adaptive fusion.
We can discover that adaptive fusion outperforms all different values of fixed $\alpha$.
These results verify the necessity of adaptive fusion of LS-term interests, and our proposed attention-based network successfully accomplishes this goal.

\vspace{0.2cm}
To conclude, we conduct extensive experiments to show the superior performance of the proposed CLSR model.
Counterfactual evaluations demonstrate that LS-term interests are successfully disentangled.
More experimental results are left in Section \ref{app::ablation}.

%% file: 4.related.tex
\section{Related Work}\label{sec::related}

\noindent\textbf{LS-Term Interests Modeling in Recommendation.} 
Traditional Markov chains-based methods \cite{rendle2010factorizing} and advanced deep learning models \cite{hidasi2015session, zhu2017next, tang2018personalized, zhou2019deep, kang2018self, sun2019bert4rec,li2017neural,li2020time,ma2020memory} fail to distinguish between LS-term interests, since a unified representation is insufficient to fully capture user interests. 
Therefore, several methods \cite{zhao2018plastic, yu2019adaptive, an2019neural, lv2019sdm, grbovic2018real,hu2020graph} were proposed that explicitly differentiate between LS-term interests. 
For example, Zhao \textit{et al.} \cite{zhao2018plastic} use matrix factorization for long-term interests and use RNN for short-term interests. 
Yu \textit{et al.} \cite{yu2019adaptive} develop a variant of LSTM for short-term interests and adopt asymmetric SVD \cite{koren2008factorization} for long-term interests. 
However, disentanglement of LS-term interests is not guaranteed since these approaches impose no supervision on the learned interests representations.
Unlike existing unsupervised approaches, we propose a self-supervised method that attains stronger disentanglement of long and short-term interests.

\noindent\textbf{Self-supervised Learning in Recommendation.}
Self-supervised learning \cite{chen2020simple, he2020momentum, chen2020improved, caron2020unsupervised, grill2020bootstrap} was recently adopted by several recommendation algorithms \cite{ma2020disentangled, zhou2020s3, xin2020self,xie2020contrastive}.
For example, Zhou \textit{et al.} \cite{zhou2020s3} developed a self-supervised sequential recommender based on mutual information maximization.
And Ma \textit{et al.} \cite{ma2020disentangled} proposed to supervise sequential encoders with latent intention prototypes.
However, those methods ignore the differences between long and short-term interests, which are crucial for accurate recommendation.
In our paper, we design a self-supervised learning method to disentangle long and short-term interests for recommendation.

\noindent\textbf{Disentanglement in Recommendation.}
Disentangled representation learning in recommendation is largely unexplored until recently \cite{wang2020disentangled,wang2021learning,ma2019learning,zheng2021disentangling}.
Ma \textit{et al.} \cite{ma2019learning} propose to learn users' multiple preferences based on Variational Auto-Encoders.
Wang \textit{et al.} \cite{wang2021learning} leverage Knowledge Graph to learn different user intentions and regularize them to be differ from each other.
However, most of these works fail to impose specific semantics to the learned multiple representations because of lacking labeled data, \textit{i.e.} unsupervised disentanglement, which has been shown to be ineffective \cite{locatello2019challenging}.
In this paper, we propose to disentangle with self-supervision by designing contrastive tasks between the learned representations and interest proxies extracted from the original interaction sequences.

%% file: 5.conclusion.tex
\vspace{-0.2cm}
\section{Conclusion and Future Work}\label{sec::conclusion}

In this paper, we propose to disentangle long and short-term interests for recommendation with a contrastive learning framework, CLSR. 
Extensive experiments and counterfactual evaluations on two large-scale datasets demonstrate that CLSR consistently outperforms SOTA baselines with significant improvements.
More importantly, we empirically show that unsupervised LS-term interests modeling can easily entangle the two aspects and lead to even poorer performance.
With the help of self-supervision, CLSR can effectively disentangle LS-term interests and achieve much better performance.
As for future work, CLSR can be easily extended since it is a highly general framework,
For example, other designs of encoders or proxies can be explored.
Deploying the proposed method to industrial systems is another important future work.

%% file: 6.appendix.tex
\section{APPENDIX}

\subsection{Datasets}\label{app::dataset}
We use two datasets to conduct experiments, including a public e-commerce dataset and an industrial short-video dataset, which are also adopted by the SOTA sequential recommendation model, SURGE \cite{chang2021sequential}.
Both of them are in million scale and collected from real-world applications.

The details of the adopted datasets are introduced as follows,
\begin{itemize}[leftmargin=*]
	\item \textbf{Taobao}\footnote{\url{https://tianchi.aliyun.com/dataset/dataDetail?dataId=649}}. This dataset \cite{zhu2018learning} is collected from the largest e-commerce platform in China, and it is widely used as a benchmark dataset for recommendation research \cite{zhu2018learning, zhu2019joint, pi2019practice}. It contains the user behaviors, including click, cart, and purchase from November 25 to December 3, 2017. We use the click data and adopt 10-core settings to filter out inactive entities. To evaluate the recommendation performance, we use all the instances till December 1 as training data. We use the instances on December 2 for validation and evaluate the final performance with the instances on December 3.
	\item \textbf{Kuaishou\footnote{\url{https://www.kuaishou.com}}.} This industrial dataset is collected from Kuaishou APP, one of the largest short-video platforms in China. Users can browse short videos uploaded by other users. We extract a subset of the logs from October 22 to October 28, 2020. The dataset contains user interactions with short videos, including click, like, follow (subscribe), and forward. We use the click data and also adopt 10-core settings to guarantee data quality. We keep the instances of the first 6 days as training set, and reserve the last day for validation (before 12 pm) and test (after 12 pm).
\end{itemize}
Table \ref{tab::dataset_appendix} shows the statistics of the two datasets after splitting.

\subsection{Baselines}\label{app::baseline}
We compare the proposed approach with the following competitive recommenders:
\begin{itemize}[leftmargin=*]
    \item \textbf{NCF} \cite{he2017neural}: This method is the state-of-the-art general recommender which combines matrix factorization and multi-layer perceptrons to capture the non-linearity of user interactions.
	\item \textbf{DIN} \cite{zhou2018deep}: This method uses attention mechanism to aggregate the historical interaction sequences. Attention weights are computed according to the target item.
	\item \textbf{LightGCN} \cite{he2020lightgcn}: This method is the state-of-the-art GCN based recommender and it utilizes neighborhood aggregation to capture the collaborative filtering effect.
	\item \textbf{Caser} \cite{tang2018personalized}: This method regards the sequence of items as images and extract sequential patterns with a convolutional network.
	\item \textbf{GRU4REC}~\cite{hidasi2015session}: This is the first approach that applies RNN to session-based recommendation system, with modified mini-batch training and ranking loss. 
	\item \textbf{DIEN} \cite{zhou2019deep}: This method improves DIN by combining attention with GRU to model the sequential pattern of user interests, and takes interests evolution into consideration.  
	\item \textbf{SASRec} \cite{kang2018self}: This method is the state-of-the-art sequential recommendation model which utilizes self-attention to capture sequential preferences.
	\item \textbf{SURGE} \cite{chang2021sequential}: This is the state-of-the-art recommendation approach which utilizes graph convolutional networks (GCN) to model user interest from sequential interactions.
	\item \textbf{SLi-Rec} \cite{yu2019adaptive}: This is the state-of-the-art algorithm which captures long-term interests with asymmetric-SVD and models short-term interests with a modified LSTM.
\end{itemize}

\begin{table}[t!]
    \caption{Statistics of the datasets.}
    \label{tab::dataset_appendix}
    \begin{tabular}{cccc}
      \toprule
      dataset & train & validation & test \\
      \midrule
      Taobao & 1,094,775 & 191,946 & 184,434 \\
      Kuaishou & 12,925,390 & 641,580 & 1,385,689 \\
      \bottomrule
    \end{tabular}
\end{table}

\subsection{Implementation Details}\label{app::implementation}
We implement all the models with the Microsoft Recommenders framework \cite{argyriou2020microsoft} based on TensorFlow \cite{abadi2016tensorflow}. We use the Adam optimizer \cite{kingma2014adam}.
Embedding size $d$ is set as 40.
We use a two-layer MLP with hidden size [100, 64] for interaction estimation.
Batch normalization is enabled for the MLP, and the activation function is ReLU.
The maximum length for user interaction sequences is 50 for Taobao dataset and 250 for Kuaishou dataset.
We use grid-search to find the best hyper-parameters.
The optimal settings for our proposed implementation are: $L_2$ regularization weight is 1e-6. Batchsize is 500. Learning rate is 0.001. $\beta$ is 0.1. $l_t$ is 5 for Taobao dataset and 10 for Kuaishou dataset. $k$ is 3 for Taobao dataset and 5 for Kuaishou dataset. $\mathcal{L}_{\text{con}}$ is $\mathcal{L}_{\text{tri}}$ for Taobao dataset and $\mathcal{L}_{\text{bpr}}$ for Kuaishou dataset.

\subsection{More Studies on the Proposed Method}\label{app::ablation}

\begin{table}[t]
    \caption{Performance of different $k$ on Taobao dataset.}
    \label{tab::taobao_k}
    \begin{tabular}{c|cccc}
      \toprule
      $k$ & AUC & GAUC & MRR & NDCG@2 \\
      \midrule
      1 & 0.8975 & 0.8927 & 0.4306 & 0.3717 \\
      2 & 0.8956 & 0.8938 & 0.4364 & 0.3798 \\
      3 & 0.8953 & 0.8936 & 0.4372 & 0.3788 \\
      4 & 0.8936 & 0.8924 & 0.4331 & 0.3747 \\
      \bottomrule
    \end{tabular}
\end{table}

In this section, we conduct experiments to investigate how the proposed method performs under different values of several introduced hyper-parameters. We also include further ablation studies on several components.

\noindent\textbf{Short-term Proxy $k$.}
In the proposed method, we use mean pooling of the recent $k$ interacted items as the proxy representation for short-term interests.
Table \ref{tab::taobao_k} illustrates the results of different $k$ on Taobao dataset.
We can observe that setting $k$ as 1 achieves poorer performance except for AUC, which means only using the last interacted item as proxy for short-term interests is not a good choice since one interaction can be noise with large possibilities.

\noindent\textbf{Interests Evolution}
Short-term interests are quite different from long-term interests with respect to their dynamics over time, thus we utilize a GRU to generate query vectors in Eqn (\ref{eq::short}) which simulates the evolution of short-term interests.
We study the effect of interests evolution and results are shown in Table \ref{tab::evolution}.
We can observe that removing interests evolution causes a significant decrease of accuracy on both datasets, which confirms the necessity of modeling different semantics of LS-term interests.

\begin{table}
\small
	\caption{Study of interests evolution.}
	\vspace{-10px}
	\label{tab::evolution}
	\begin{tabular}{c|c|ccccc}
    \toprule
    Dataset & Evolution & AUC & GAUC & MRR & NDCG@2 \\ 
    \midrule
    \multicolumn{1}{c|}{\multirow {2}{*}{Taobao}} & yes & 0.8953 & 0.8936 & 0.4372 & 0.3788 \\
    \multicolumn{1}{c|}{} & no & 0.8847 & 0.8884 & 0.4320 & 0.3735 \\
    \hline
    \multicolumn{1}{c|}{\multirow {2}{*}{Kuaishou}} & yes & 0.8563 & 0.8718 & 0.9382 & 0.9544 \\
    \multicolumn{1}{c|}{} & no & 0.8202 & 0.8333 & 0.9226 & 0.9429 \\
    \bottomrule
    \end{tabular}
\end{table}

\noindent\textbf{Study of Different Design Choices}
We further compare different design choices in CLSR.
Specifically, we investigate different options for short-term interests encoder and contrastive loss function in Eqn (\ref{eq::rho}) and (\ref{eq::L_con}).
For the RNN $\rho$ in the short-term interests encoder $\psi$, we compare LSTM \cite{hochreiter1997long}, GRU \cite{chung2014empirical}, and Time4LSTM proposed by SLi-Rec \cite{yu2019adaptive}.
For $\mathcal{L}_\text{con}$, we compare BPR loss and triplet loss.
Table \ref{tab::design} shows the results of different design choices.
We can observe that Time4LSTM outperforms LSTM and GRU on both datasets, indicating that the time interval feature is helpful for LS-term interests modeling, which is ignored by LSTM and GRU.
As for contrastive loss, each loss function fails to consistently outperform the competitor, which can be explained by the different scales of the two datasets.
In fact, CLSR is a highly general framework in which many sequential encoders and loss functions can be utilized. 
We leave the further study as future work.

\begin{table}
\small
	\caption{Comparison of different design choices.}
	\vspace{-10px}
	\label{tab::design}
	\begin{tabular}{c|ccc|cc}
    \toprule
    Dataset & LSTM & GRU & Time4LSTM & BPR & Triplet \\ 
    \midrule
    Taobao & 0.8872 & 0.8860 & \bf{0.8953} & 0.8909 & \bf{0.8953} \\
    Kuaishou & 0.8240 & 0.8259 & \bf{0.8563} & \bf{0.8563} & 0.8102\\
    \bottomrule
    \end{tabular}
    \vspace{-10px}
\end{table}

\noindent\textbf{Fusion Predictor GRU.}
In the proposed adaptive fusion model based on the attention technique, we incorporate both the target item and the historical sequence to predict whether the next interaction is driven by long or short-term interests.
Specifically, we adopt a separate GRU that takes the historical sequence as input, and we use the final state as the input of MLP.
We conduct experiments to investigate whether taking the historical sequence into consideration is necessary.
Table \ref{tab::fusion_gru} illustrates the results of the proposed method with and without the fusion predictor GRU.
We can observe that removing the fusion predictor GRU makes the recommendation performance drop significantly, which confirms that the importance of long or short-term interests is largely determined by the historical sequence.

\begin{table}[t]
	\caption{Study of fusion predictor GRU on Taobao dataset.}
	\label{tab::fusion_gru}
	\begin{tabular}{c|ccccc}
    \toprule
    Method & AUC & GAUC & MRR & NDCG@2 \\ 
    \midrule
    w/ GRU & 0.8953 & 0.8936 & 0.4372 & 0.3788 \\
    w/o GRU & 0.8817 & 0.8853 & 0.4275 & 0.3692 \\
    \bottomrule
    \end{tabular}
\end{table}

\noindent\textbf{Attentive Encoder}
As introduced in Equation (\ref{eq::att1}), the proposed attentive encoder adopts a MLP to compute attention weights.
The inputs of the MLP are composed of the key vector, query vector, the element-wise difference, and multiplication of key and query.
We compare the MLP based attention with simple inner product based attention:
\begin{equation}
    \alpha_k'= \langle\bm{v_k}, \bm{u_l}\rangle.
\end{equation}
Table \ref{tab::attention} illustrates the comparison of MLP and inner product based attention.
Results in Table \ref{tab::attention} show that MLP based attention outperforms inner product-based attention with a significant margin, which indicates that the relation between user interests and historical sequences is non-linear and can not be well captured by linear operators like the inner product.

\begin{table}[t]
	\caption{Study of attentive encoder on Taobao dataset.}
	\label{tab::attention}
	\begin{tabular}{c|ccccc}
    \toprule
    Attention & AUC & GAUC & MRR & NDCG@2 \\ 
    \midrule
    MLP & 0.8953 & 0.8936 & 0.4372 & 0.3788 \\
    Inner Product & 0.8684 & 0.8706 & 0.4051 & 0.3480 \\
    \bottomrule
    \end{tabular}
\end{table}
\noindent\textbf{Discrepancy Supervision}
In our proposed method, we do not add an extra discrepancy loss on the LS-term interests to make them independent with each other as other works \cite{wang2020disentangled,zheng2021disentangling}, since we believe self-supervision is enough to accomplish disentanglement.
During our experiments, we tried to add an independent loss between the two interests as, and AUC drops by 0.01, which verifies our point.
It is worthwhile to notice that many existing works \cite{ma2020disentangled,cen2020controllable,wang2020disenhan} also did not use the independent loss.

\noindent\textbf{More Counterfactual Evaluations}
\begin{figure}[t]
  \centering
  \includegraphics[width=\linewidth]{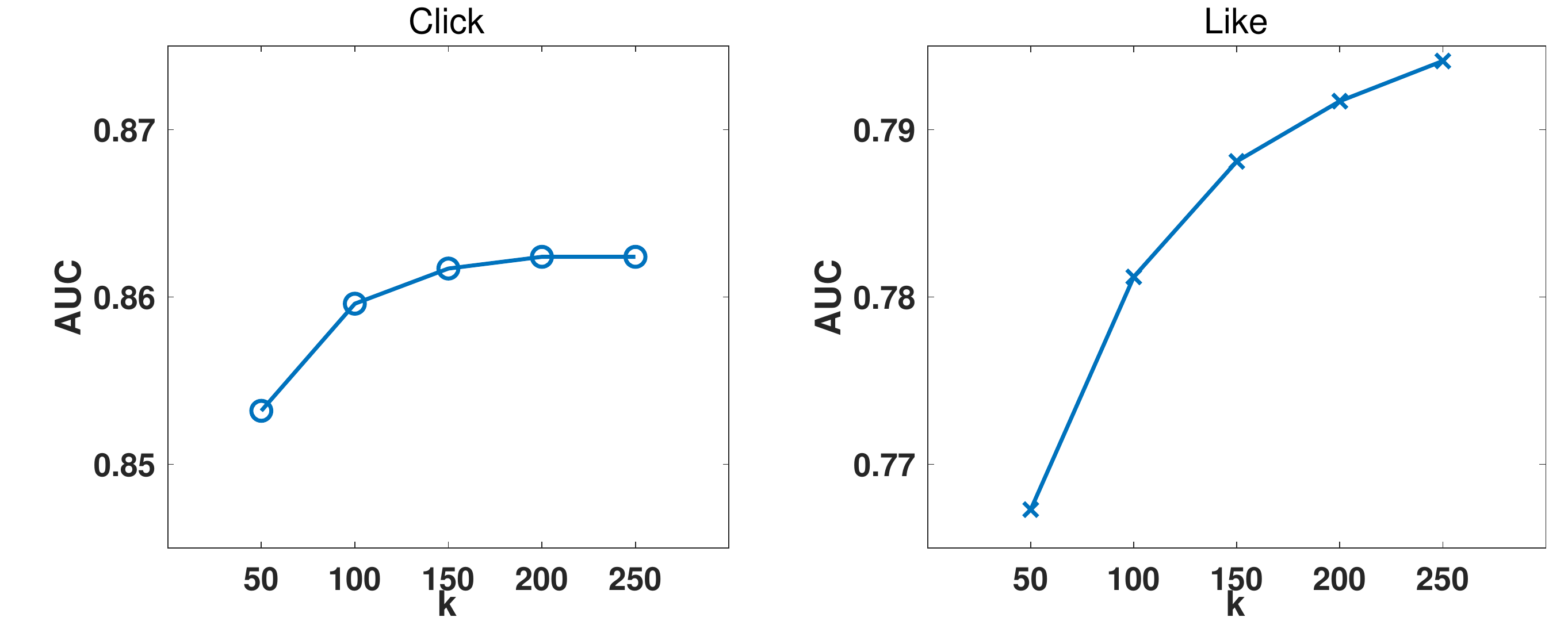}
  \vspace{-15px}
  \caption{Counterfactual (truncate) evaluation of the proposed method on Kuaisohu dataset.}
  \Description{Counterfactual (truncate) evaluation of the proposed method on Kuaisohu dataset.}
  \label{fig::truncate_kuaishou}
\end{figure}
Figure \ref{fig::truncate_kuaishou} illustrates the AUC of click and like on Kuaishou dataset with available history length $k$ varying from 50 to 250.
Results on Kuaishou dataset are in line with results on Taobao dataset in Figure \ref{fig::truncate}(a).
Specifically, AUC of like is more sensitive to the length of available history and improves drastically as $k$ increases, while AUC of click does not improve much as we increase $k$.
Since like reflects more about the user's long-term interest, it is necessary to have access to the entire user interaction sequence.
Meanwhile, click is more about short-term interest, and thus it can be largely captured from the recent history, and looking back to early history will not bring further gains.

\noindent\textbf{Complexity.}
We use a single GPU to compare the complexity.
The training time of CLSR and typical baselines on Taobao dataset are shown in Table \ref{tab::complexity}.
The parameter scale of CLSR is comparable with SLi-Rec (both takes 4.1Gb GPU memory).

\begin{table}[t!]
	\caption{Training time cost on Taobao dataset.}
	\label{tab::complexity}
	\begin{tabular}{c|ccc}
    \toprule
    Method & GRU4REC & SLi-Rec & CLSR \\ 
    \hline
    Time &  27.8min & 26.7min & 28.2min \\
    \bottomrule
    \end{tabular}
\end{table}